# Tensile-strained germanium-on-insulator substrate fabrication for silicon-compatible optoelectronics

J. Raja Jain,[1,*] Dany-Sebastien Ly-Gagnon,[1] Krishna C. Balram,[1] Justin S. White,[2] Mark L. Brongersma,[3] David A. B. Miller,[1] and Roger T. Howe[1]

[1]*Department of Electrical Engineering, Stanford University, Stanford, CA 94305, USA*
[2]*Department of Applied Physics, Stanford University, Stanford, CA 94305, USA*
[3]*Department of Materials Science and Engineering, Stanford University, Stanford, CA 94305, USA*
[*]*jrjain@stanford.edu*

**Abstract:** We present a method to fabricate tensile-strained germanium-on-insulator (GOI) substrates using heteroepitaxy and layer transfer techniques. The motivation is to obtain a high-quality wafer-scale GOI platform suitable for silicon-compatible optoelectronic device fabrication. Crystal quality is assessed using X-Ray Diffraction (XRD) and Transmission Electron Microscopy. A biaxial tensile film strain of 0.16% is verified by XRD. Suitability for device manufacturing is demonstrated through fabrication and characterization of metal–semiconductor–metal photodetectors that exhibit photoresponse beyond 1.55 μm. The substrate fabrication process is compatible with complementary metal–oxide–semiconductor manufacturing and represents a potential route to wafer-scale integration of silicon-compatible optoelectronics.

## References and links

## 1. Introduction

Optoelectronic devices fabricated using III–V materials can achieve high performance due to the presence of a direct bandgap. However, these materials have the disadvantages of high cost, high process complexity, and incompatibility with silicon (Si) complementary metal–oxide–semiconductor (CMOS) manufacturing. Recently, germanium (Ge) has been investigated by several groups as a potential material platform for various optoelectronic components [1–3]. Its inherent compatibility with Si–CMOS processing makes it attractive for integrating optoelectronics with conventional electronics. Its absorption coefficient in the 1.3–1.47 µm wavelength range is comparable to those of some III–Vs. In addition, the direct Ge Γ-valley is only 136 meV higher in energy than the indirect L-valley and lies within the telecommunications C-band, suggesting that Ge has potential as a useful light emitter [4,5].

The majority of Ge-based optoelectronic devices in the literature use thick Ge films epitaxially grown on Si [1,6]. Thick epitaxial Ge layers have several disadvantages, including poor device isolation due to the direct heteroepitaxial growth on Si, high defect concentrations at the Ge/Si interface, and larger device capacitance due to film thickness requirements. The germanium-on-insulator (GOI) material system represents a more attractive platform for Si-compatible optoelectronics. To this end, several groups have developed processes for fabricating GOI films, including wafer-scale approaches, such as condensation and smart-cut [2,7–11], as well as local approaches, such as zone melting and rapid melt growth [3,12–14]. The wafer-scale approaches are more attractive from a manufacturing perspective; however, current methods require tight process control and limit the range of thicknesses that can be obtained in the final GOI film.

In this letter, we present a method of obtaining tensile-strained wafer-scale GOI films using Ge-on-Si heteroepitaxy and layer transfer techniques. Film quality is investigated using X-Ray Diffraction (XRD) and Transmission Electron Microscopy (TEM), while film strain is examined via XRD analysis. Photocurrent absorption measurements of metal–semiconductor–metal (MSM) photodetectors fabricated on these films show that the bandgap shift is preserved and that high-quality photodetectors covering the telecommunications C-band could be fabricated using this material platform. These material and device studies suggest that the demonstrated tensile-strained GOI fabrication method can facilitate the production of Si-compatible, high-performance Ge-based optoelectronics by a controllable, wafer-scale process.

## 2. Experimental methods

Approximately 1.57 µm of Ge was epitaxially grown by Reduced Pressure Chemical Vapor Deposition (RPCVD) on standard 525-µm-thick Si substrates (seed wafers) using the Multiple

Hydrogen Annealing for Heteroepitaxy (MHAH) technique [15]. Thermal SiO$_2$ was grown on Si handle wafers at 1000°C using an atmospheric furnace with wet ambient. After a thorough rinse/dry process, seed and handle wafers were vacuum bonded at 50°C and 0.075 mTorr with 1 bar applied bonding pressure. Bonded wafer pairs were then annealed for 10 hrs at 800°C in a nitrogen-rich ambient at atmospheric pressure. Figure 1a presents a cross-section TEM image of a GOI stack with a 1-µm-thick buried oxide, while Figure 1b shows a high-resolution cross-section TEM image of the Ge/SiO$_2$ bonding interface. The Ge/SiO$_2$ interface is smooth and uniform across the wafer, while the epitaxial Ge film has high crystal quality as evidenced by the selective area diffraction (SAD) pattern (inset).

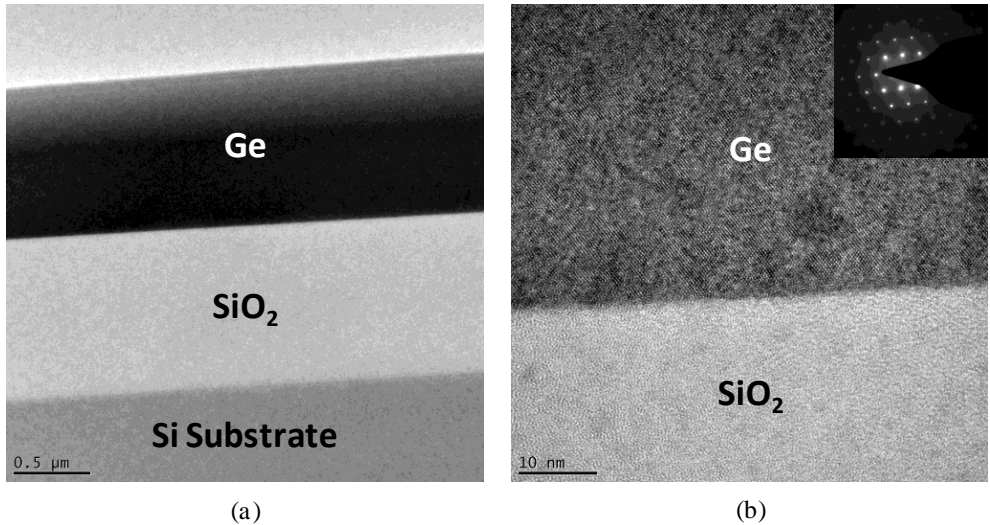

Figure 1. Cross-section images of GOI material. (a) Cross-section TEM image of a representative GOI stack. (b) Cross-section high-resolution TEM image of Ge/SiO$_2$ bonding interface (inset: selective area diffraction (SAD) pattern indicating high-quality, single-crystal Ge).

After the post-bond anneal, the Si seed wafers were commercially back-ground, leaving ~50 µm of Si as a buffer layer on top of Ge. The Si buffers were then chemically removed by exposing the bonded wafer pairs to a 25% tetramethyl ammonium hydroxide (TMAH) solution at 95°C. No measureable etching or damage of the Ge surface was observed.

Although the MHAH process yields a relatively defect-free Ge surface, the Ge/Si epitaxial interface has a high concentration of crystal defects and dislocations. In the above wafer bonding process, this epitaxial interface results in a defect-rich GOI surface after the layer transfer and Si seed wafer removal steps are complete. In order to obtain a high-quality, low–defect density GOI film suitable for device fabrication, the defective surface layer must be removed. A chemical–mechanical polishing (CMP) system with colloidal silica slurry was used to thin the Ge film on two different samples, hereafter referred to as GOI1 and GOI2, each having a 1-µm-thick buried oxide. In order to enhance the polishing rate, a small amount of hydrogen peroxide (~0.2%) was added to the slurry mixture [16]. Final GOI film thicknesses for GOI1 and GOI2 were 1.34 µm and 0.19 µm, respectively.

Figure 2 presents triple-axis XRD scans of the GOI films (a) before CMP treatment and for (b) GOI1 and (c) GOI2. The as-bonded Ge XRD peak shown in Figure 2a indicates a defect-rich film with some Si intermixing at the top surface as a likely result of the epitaxial growth and anneal processes. The increasingly sharper Ge XRD peaks for GOI1 and GOI2 indicate that, as the GOI layer is progressively thinned, regions with higher defect densities are removed and average film quality improves. Strain analysis of the XRD scans indicates the presence of 0.16% biaxial tensile strain in the final GOI films.

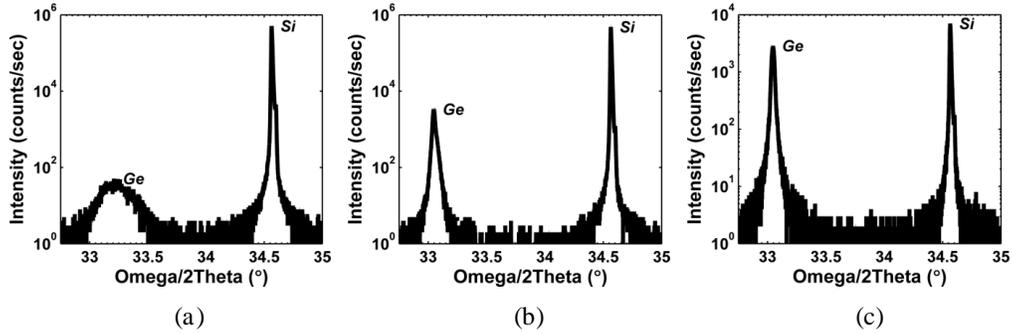

Figure 2. Triple-axis XRD scans indicating improvement in average film quality with CMP treatment. (a) As-bonded GOI wafer (film thickness = 1.57 μm). (b) GOI1 (film thickness = 1.34 μm; 0.16% biaxial tensile strain). (c) GOI2 (film thickness = 0.19 μm; 0.16% biaxial tensile strain).

Figure 3 shows representative plan-view TEM images of the Ge film surface for (a) GOI1 and (b) GOI2. Consistent with the XRD results, plan-view TEM analysis indicates that progressive thinning and removal of the Ge surface leaves GOI films with smaller average defect densities. Quantitative inspection of multiple plan-view images per sample suggests defect densities of $< 3 \times 10^8$ cm$^{-2}$ (range: $1.5 \times 10^8$ cm$^{-2}$–$4.5 \times 10^8$ cm$^{-2}$) for GOI1 and $< 5 \times 10^7$ cm$^{-2}$ (range: $1 \times 10^7$ cm$^{-2}$–$1 \times 10^8$ cm$^{-2}$) for GOI2.

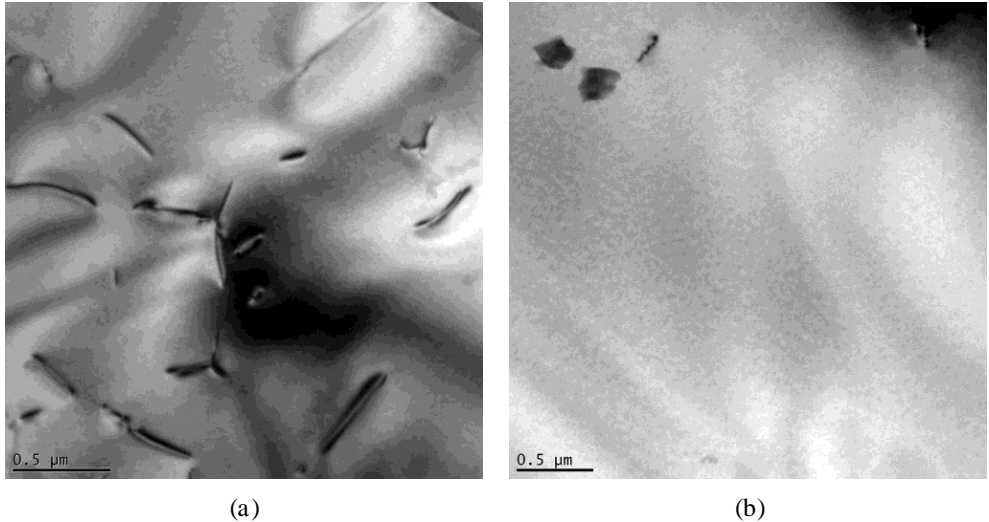

Figure 3. Plan-view TEM images indicating reduction in average defect density with CMP treatment. (a) GOI1 (average defect density $< 3 \times 10^8$ cm$^{-2}$). (b) GOI2 (average defect density $< 5.5 \times 10^7$ cm$^{-2}$).

## 3. Device fabrication and characterization

In order to examine the shift in the Ge band edge due to the tensile strain present in the film, MSM photodetectors were fabricated on GOI1 and GOI2. 70 nm Au/5 nm Ti contact pads were e-beam evaporated on the tensile-strained GOI films and patterned via photoresist lift-off to leave electrodes with an interdigitated finger width × spacing of 5 μm × 5 μm.

Devices were illuminated using an optical signal from a broad-spectrum Fianium SC450-2 laser source filtered by a SpectraPro 2300i monochromator chopped at 1 kHz and focused with a long–working distance 20× objective (numerical aperture of 0.4). Collected photocurrent (1V bias) was amplified with a Stanford Research Systems SR570 preamplifier before being fed into a Stanford Research Systems SR870 lock-in amplifier. The incident optical power measured at the device position was used to calculate normalized photocurrent.

Figure 4 summarizes the characterization results of MSM photodetectors built on GOI1 (thick GOI) and GOI2 (thin GOI). Both samples exhibit a reduction in the Ge direct bandgap energy and a shift towards longer wavelengths due to the biaxial tensile strain present in the GOI layers. Periodic peaks in the photodetection spectra are the result of Fabry–Perot reflections from the GOI material stacks. Additionally, the device fabricated on GOI2 exhibits a ~2× enhancement in device photoresponse compared to GOI1, including a flatter photodetection response across wavelengths (partially due to a change in the Fabry–Perot behavior). These results indicate a lower average defect density and an improved overall film quality for the thinner GOI2 device layer. The Ge film strain and absorption edge exhibited by both films are consistent with previously reported results [1]. The longer wavelength cut-off indicates that strain is preserved in the bonded stacks and demonstrates that this GOI substrate fabrication method can produce an attractive platform for fabrication of strained Ge films on insulator.

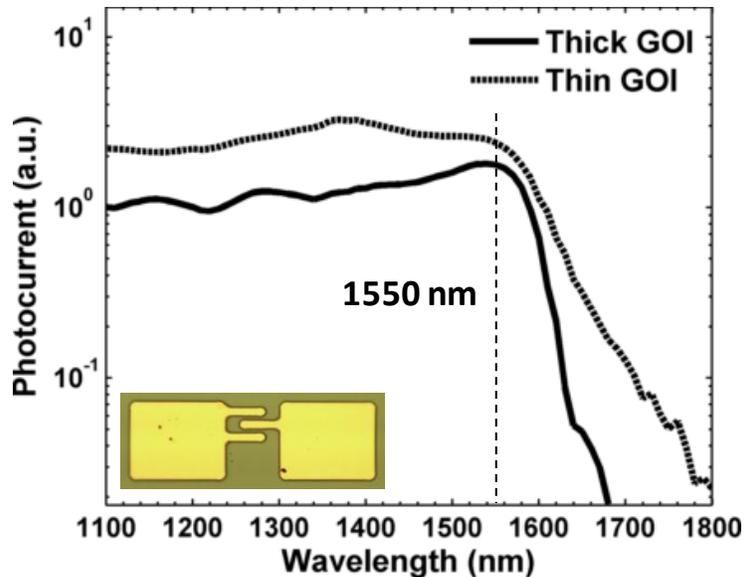

Figure 4. Characterization results of MSM photodetectors built on GOI1 (thick GOI) and GOI2 (thin GOI). Bulk Ge has a direct bandgap energy of 0.8 eV (1550 nm).

## 4. Conclusion

A method to fabricate low defect density, biaxial tensile-strained GOI substrates using Ge heteroepitaxy and layer transfer techniques has been demonstrated. Crystal quality and film strain have been studied using XRD and TEM analyses. The absorption spectra of MSM photodetectors fabricated on the GOI substrates demonstrate that this material could be used for detection of optical signals in the telecommunications C-band. The tensile-strained GOI fabrication approach represents a potential route to Si-compatible optoelectronics integration at the wafer scale.

As an example implementation, a tensile-strained Ge optoelectronic device layer can be added to a conventional CMOS electronics manufacturing process by using the described GOI fabrication approach. In this implementation, a low-temperature oxide can be deposited after the fabrication of conventional CMOS electronics on a Si handle wafer. Following a planarization step and surface pre-treatment, the tensile-strained Ge layer can be transferred to the Si handle wafer. Vias and metal layers can subsequently be used to interconnect the CMOS circuitry and Ge device layer to produce an integrated optoelectronic chip architecture.

**Acknowledgements**

This work was done in the Stanford Nanofabrication Facility of the National Nanotechnology Infrastructure Network. The authors would like to thank Marika Gunji of Stanford University for help with TEM imaging. Financial support was provided by the MARCO Interconnect Focus Center and a Stanford University Center for Integrated Systems gift grant.